# IMPROVED ALGORITHMS FOR NANOPORE SIGNAL PROCESSING

N. Arjmandi[1, 2], W. Van Roy[1], L. Lagae[1, 2], G. Borghs[1, 2]

[1] IMEC, Kapeldreef 75, 3001 Leuven, Belgium
[2] Department of Physics and Astronomy, KU Leuven, Celestijnenlaan, 200D, 3001 Leuven, Belgium

## 1 INTRODUCTION

Nanopores are promising devices for detection and characterization of nanometer-sized analytes suspended in a liquid; these include nanoparticles [2], DNA [3], RNA [4], viruses [2] and proteins [5]. These devices are basically a nanometer-sized pore that connects two microfluidic chambers. Translocation of analytes through the pore introduces a temporal change in its resistance that can be recorded as a spike in the ionic current that passes through the pore or a voltage spike. All applications of nanopore devices are based on detection of translocation spikes in the recorded signal and extraction of amplitude, duration and their rate of occurrence [2-5]. The recorded signal is usually considerably noisy [6], with a significant baseline drift [6] and more than hundreds of translocation spikes that may vary in shape and size [2]. Thus, incorporation of suitable signal processing algorithms is necessary for correct and fast detection of all the translocation spikes and accurate measurement of their amplitude and duration. Nanopore devices are subjected to intense research and significant improvements; however, there are only a few reported investigations on processing the output signals of these devices [7-14]; while these methods are determinant of the technique's precision, accuracy and applicability.
Nanopore signal processing generally consists of baseline removal, denoising, spike detection, extraction of spikes' amplitude and duration, and classification of spikes. Here we present an improved method for baseline removing, an optimized algorithm for denoising the nanopore signals, a novel method for spike detection that detects all the translocation spikes more correctly, and an improved and physically meaningful algorithm for measuring the duration and amplitude of the translocation spikes [1].

## 2 METHOD AND RESULTS

### 2.1 Experimental Results

The proposed software has been used to process ionic current signals of a solid-state nanopore (fig. 1) using a range of different nanoparticles. The obtained results are compared with the results of the conventional software (Clampfit, MDS Analytical Technologies) in addition to results of a commercially available dynamic light scattering (DLS) system. In order to fabricate the nanopore, silicon on insulator wafers were coated by nitride. Electron beam lithography [15] was used for patterning a square on the front side and KOH was used for etching a pyramidal pit into the silicon on the front side. Then, a pyramidal pit etched on the back side to open the nanopore's end, and the nanopore was packaged between two microfluidic chambers as described elsewhere [16]. Citrated gold nanoparticles (Ted Pella Inc.) were diluted ten times in 300mM KCl solutions, and put on the front side of the nanopore device. 120 mV was applied between the two sides of the nanopore and the ionic current recorded by an Axopatch 200B patch clamp amplifier. A MiniDigi (Molecular Devices) was used for analog to digital conversion and pCLAMP 10 (Molecular Devices) was used for data acquisition.

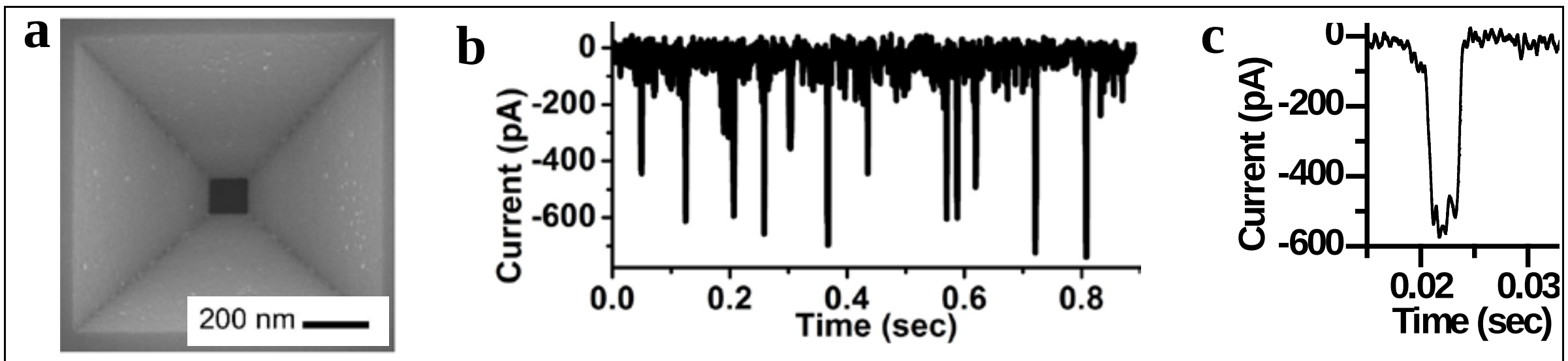

**Figure 1** *The nanopore experiment. a) Top view of a 120 nm wide pyramidal nanopore. b) A recorded ionic signal resulting from a mixture of particles that is mentioned in the text. c) Close-up representation of a typical translocation spike.*

Different types of particles result in different amplitudes (fig. 2.a, d). We have obtained narrower distributions, more precise and better-separated populations, using the proposed software comparing to the conventional methods. To present these properties in a practical experiment, we have used a mixture of 20, 40, 50, 60, and 80 nm nanoparticles and used different methods to process the recorded signal. Conventional methods do not result in well separate populations at the expected points (fig. 2.b), while the proposed software results in more accurate and separable populations (fig. 2.c). The origin of such an improvement and the details of the proposed methods are described hereunder.

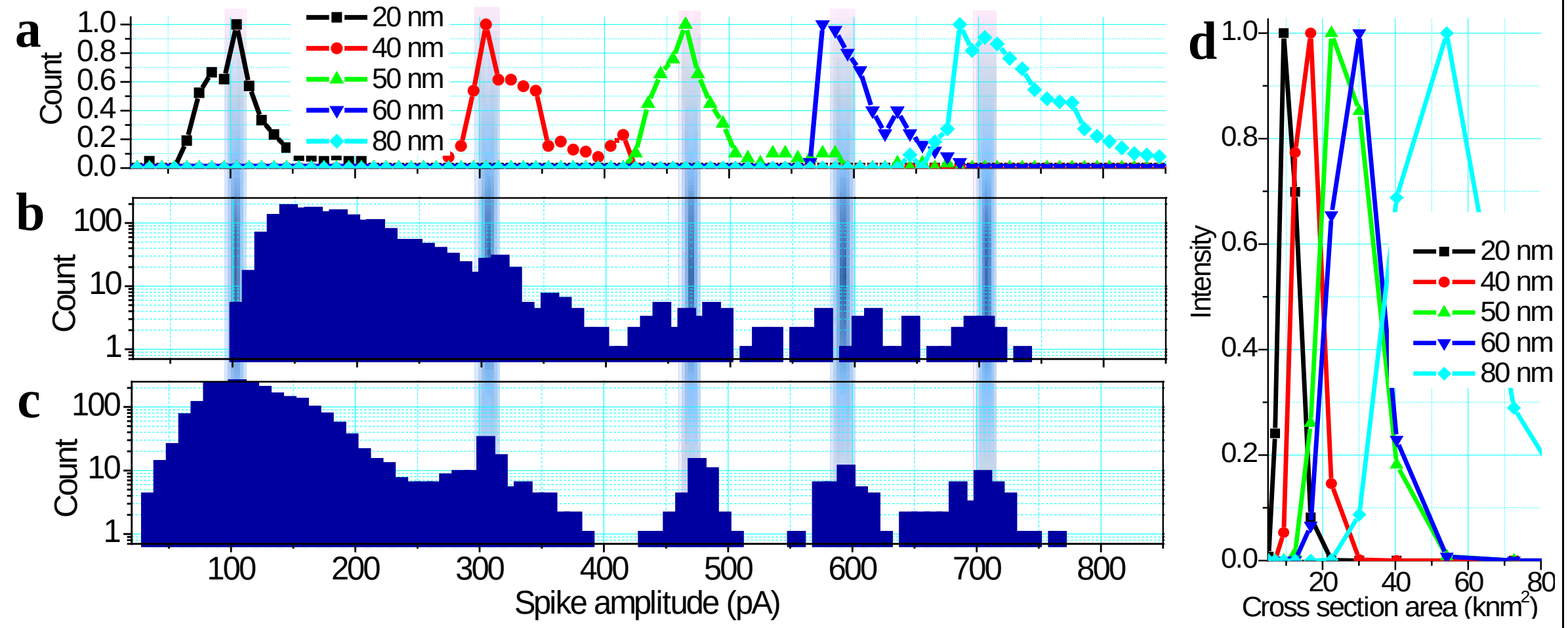

**Figure 2** *Comparing the conventional and the proposed methods. a) Normalized histogram of spike's amplitudes obtained by the proposed method from separated nanopore measurements using 20, 40, 50, 60 and 80 nm gold nanoparticles. b) Histogram resulted from the conventional thresholding method from mixture of the aforementioned particles. c) Histogram obtained from the proposed method from the same signal. d) Separated DLS measurements results of the particles.*

### 2.2 Baseline Removing

Although, the proposed software measures the amplitude and duration of the translocation spikes independent of signal's baseline, it first removes the baseline to enhance the visualization of the signals in addition to extraction of the equivalent circuit parameters of the device. The baseline drift mainly originates from the membrane as a lossy capacitance [6]. The most widely used method for baseline removing is the so called moving window (MW), which is averaging a number of data points around each translocation spike [2,8,14]. These algorithms are prone to significant inaccuracies in baseline detection in some frequent happening conditions (fig. 3). Consideration of the membrane as a lossy capacitor [6], the nanopore as a resistor, bulk liquid as a resistor, and electrodes as electric double layer capacitors in parallel with tunneling resistors; results in a summation of infinite number of exponential for the step response of the nanopore. However it is sufficiently precise to consider the first two exponential. Hereby, the baseline can be found by fitting this summation of exponentials to the upper or lower envelope of the signal. Knowing the parameters of this fitted equation, the parameters of the equivalent circuit can be calculated.

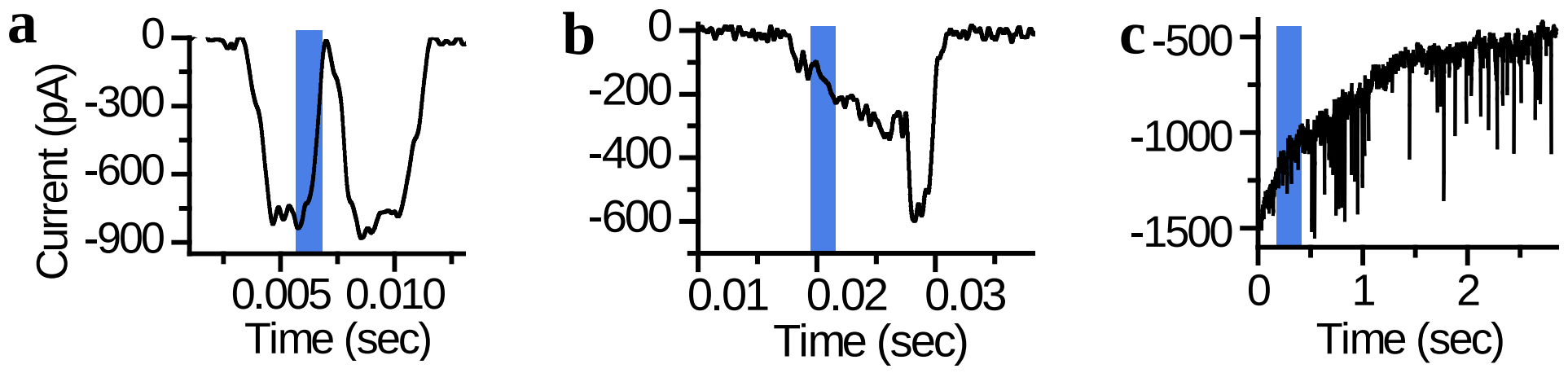

**Figure 3** *Some deficiencies of the moving window (MW) baseline detection. a) When there are two adjacent spikes, MW detects the first spikes as the others baseline. Here although the baseline is 0 pA, the MW has detected -381 pA. b) When spikes have a tail. Here although the baseline is 0 pA, MW has detected -175 pA. c) Since is not possible to reduce the window too much. MW's error increases when the baseline drifts too fast.*

### 2.3 Denoising

Low-pass filtering is generally used for denoising the nanopore signals [2, 6, 17, 18] and introduces a trade-off between signal to noise ratio and measurement band width. Due to the fact that the translocation spikes are pulse shaped [2-14] and contain a broad range of frequencies, low-pass filters can only be used at very high cut off frequencies to remove the digitization noise. In addition to using a high cut off frequency, proper choice of filter is necessary to preserve the signal. To choose the best filter, we have examined Bessel 8-pole, Gaussian, Butterworth 8-pole, Boxcar, Chebyshev 8-pole, RC 8-pole and RC 1-pole low pass filters. Although all of these filters can remove the digitization noise, the least deformation and phase shift in the signal achieved with the Gaussian.

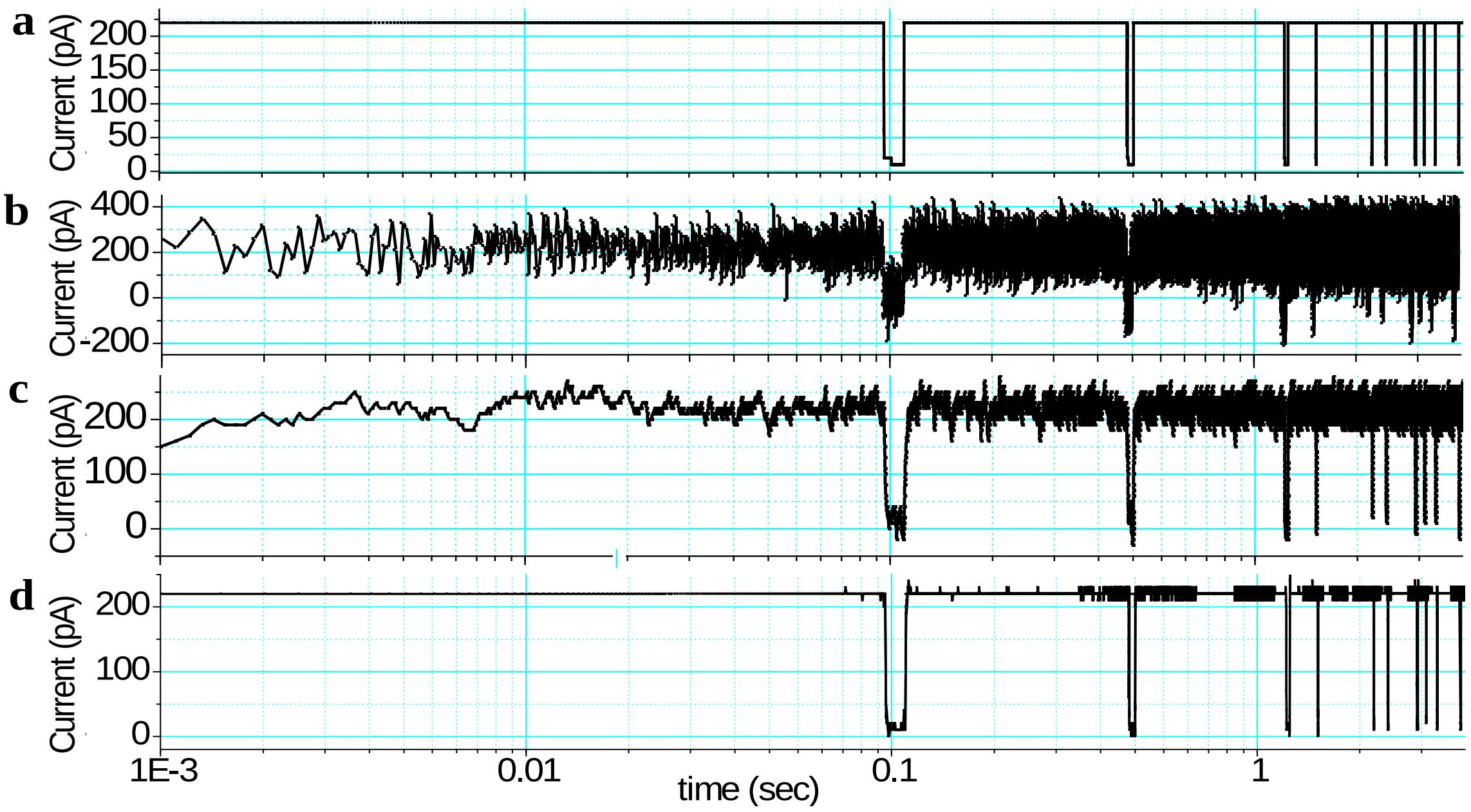

**Figure 4** *Simulated signals (time axis is scaled logarithmically). a) 10 typical spikes. b) The simulated noise added to the simulated spikes (10 kHz sampling rate). c) Low-pass filtered of the signal in b. d) The signal in c that denoised by Bior3.9.*

A proper denoising increases the detection range of the nanopore by enabling the detection of smaller analytes in a bigger nanopore, in addition to detection of smaller features of the analytes and increased precision. Jagtiani et al. have described denoising of Coulter signals by wavelets in addition to the low-pass filter [19]. These authors have optimized the choice of wavelet and wavelet thresholding level by a cross validation method [20, 21]. However this method optimizes the general shape of the denoised signal, while we are interested mostly in its amplitude and duration; furthermore, it is a probabilistic approach and does not result in certain results. Thornton et al. [7, 9, 13] have introduced wavelet denoising for nanopore signals without optimizing it. However, wavelet denoising is a strong function of type of wavelet, threshold level and signal's shape. In addition, optimization of the denoising for accurate extraction of amplitude and duration is needed. Thus, we have examined 54 different wavelets at 15 thresholding levels to choose the best denoising configuration. For this purpose, all the forces that are acting on a nanoparticle in a 120 nm nanopore have been calculated and a Monte Carlo simulation was used to calculate the nanoparticle's trajectory and the resulting spikes. The shapes of these spikes are in agreement with the experimental measurements (fig. 1.c). Ten translocation spikes were simulated (fig. 4.a), noise with the same power spectrum of the experiment is generated and added to the simulated signal (fig. 4.b); and then it was low-pass filtered as mentioned earlier. Different denoising algorithms and configurations were applied to this signal and the integral of root square error (IRSE) of the denoised signal in comparison to the originally simulated noise-less signal was calculated (fig. 5.a). About 9 different denoising configurations, which are resulting in the smallest IRSEs, were found. To choose the best of them relying on IRSE is not reliable and optimal; thus, the amplitude and duration of the denoised spikes measured and the correlation coefficients of these parameters with the originally simulated noise-less spikes have been calculated (fig. 5.b).

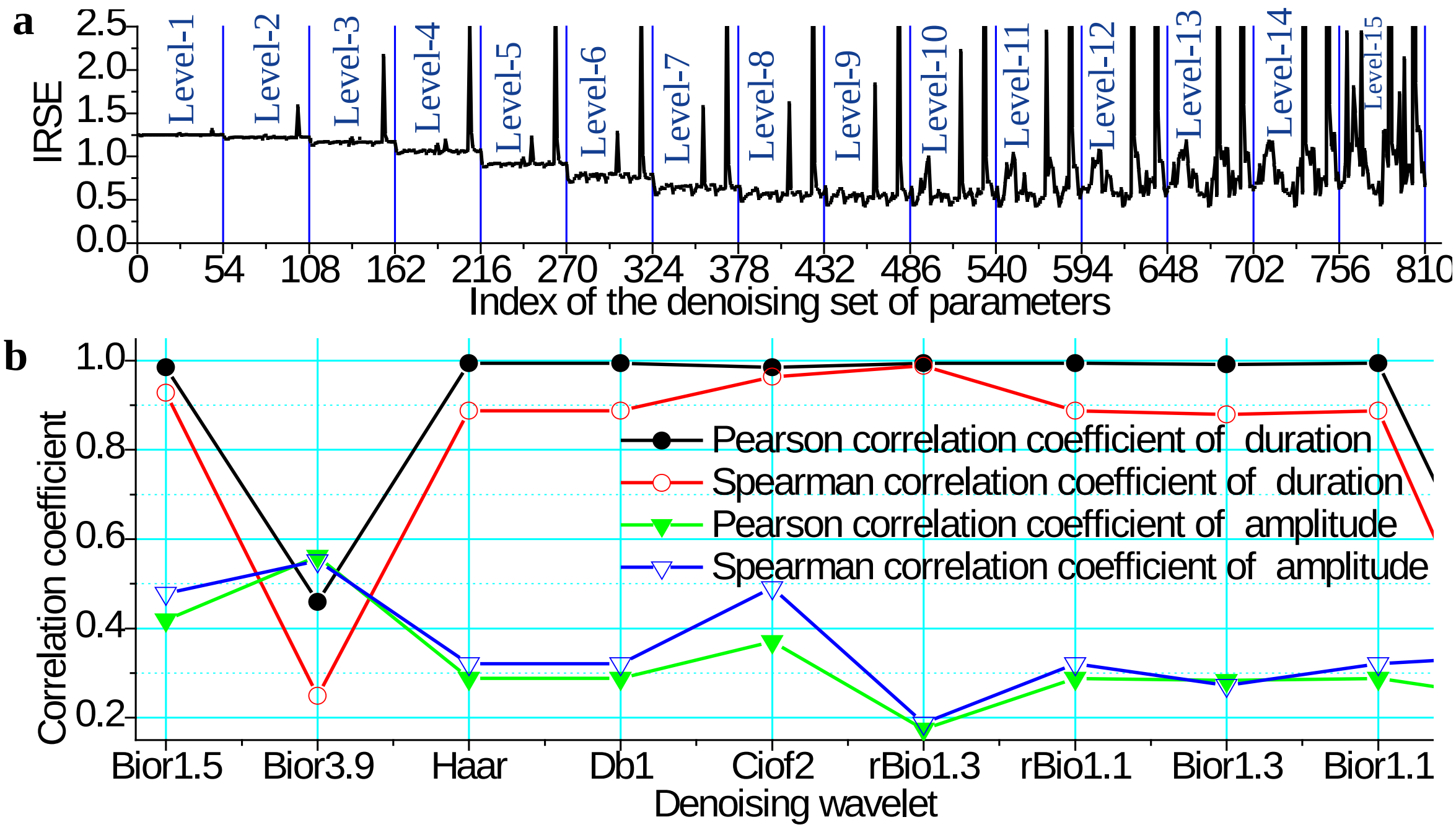

**Figure 5** *Comparing the denoising algorithms. a) IRSE of the signals that are denoised by 810 different methods and conditions. These methods are consisting of 54 different wavelets in 15 different levels. The used wavelets are: Demy, Db1 to 10, Sym2 to 8, Coif1 to 5, Bior1.1, 1.3, 1.5, 2.2, 2.4, 2.6, 2.8, 3.3, 3.5, 3.7, 3.9, 4.4, 5.5, and 6.8, rBio1.1, 1.3, 1.5, 2.2, 2.4, 2.6, 2.8, 3.3, 3.5, 3.7, 3.9, 4.4, 5.5 and 6.8. b) Comparison of the correlation coefficients of the 9 wavelets that have the lowest IRSE. Denoising by Bior3.9, results in the highest correlation between the amplitudes of the noise-less signal and the denoised signal. Denoising by rBio1.3 is relatively better for duration measurement and it results in almost ideally precise measurement of translocation duration.*

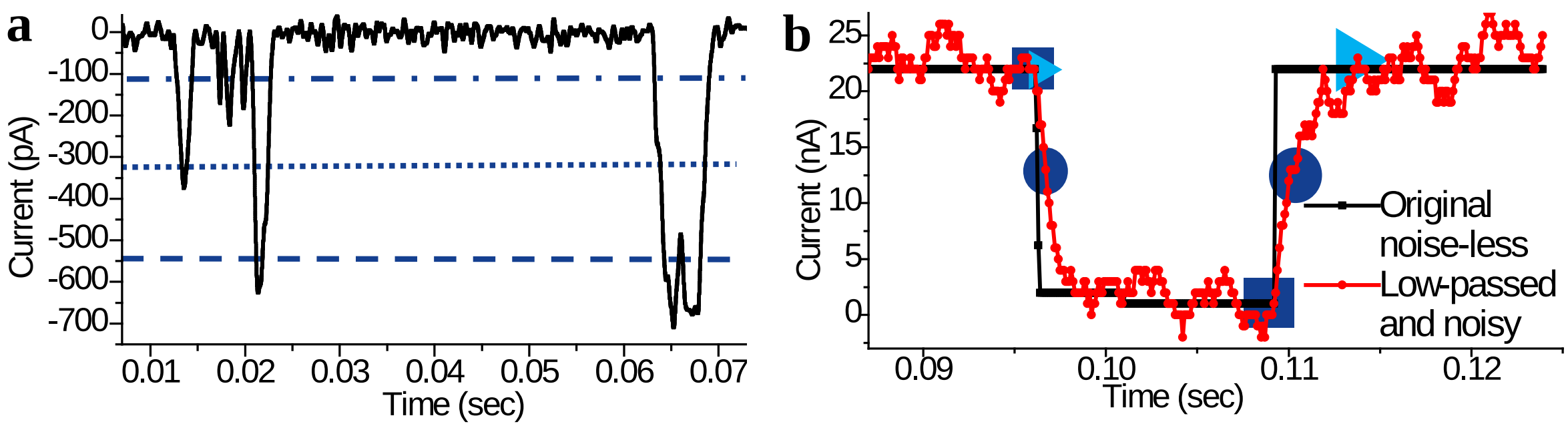

**Figure 6** *Spike analysis. a) Every threshold we choose, some of the spikes that are not crossing the threshold will not be detected. b) Starting and ending points that determined by the proposed algorithm (■) and thresholding (baseline:►, full width half max:●).*

## 2.4 Spike Detection

The current method for spike detection is the so called thresholding [2-14, 22]. In this method, a horizontal line is drawn about 5 times of noise level away from the base line; and when the signal crosses this line, algorithm reports a spike; fluctuations which are not crossing the threshold are left undetected (Fig. 6.a).

To solve this problem we have developed a novel algorithm, which detects any fluctuation in the signal which is physically meaningful – i.e. it is not probable to be originated from the electrical noise. This algorithm can detect the fluctuations within a spike, in addition to differentiation of adjacent spikes. To do so, it extracts all the local minima and maxima of the signal and considers the adjacent extrema with amplitude difference more than five times the noise level. Then, a selection algorithm finds the spikes and extracts their information. Block-diagram of the whole software is depicted in figure 7.

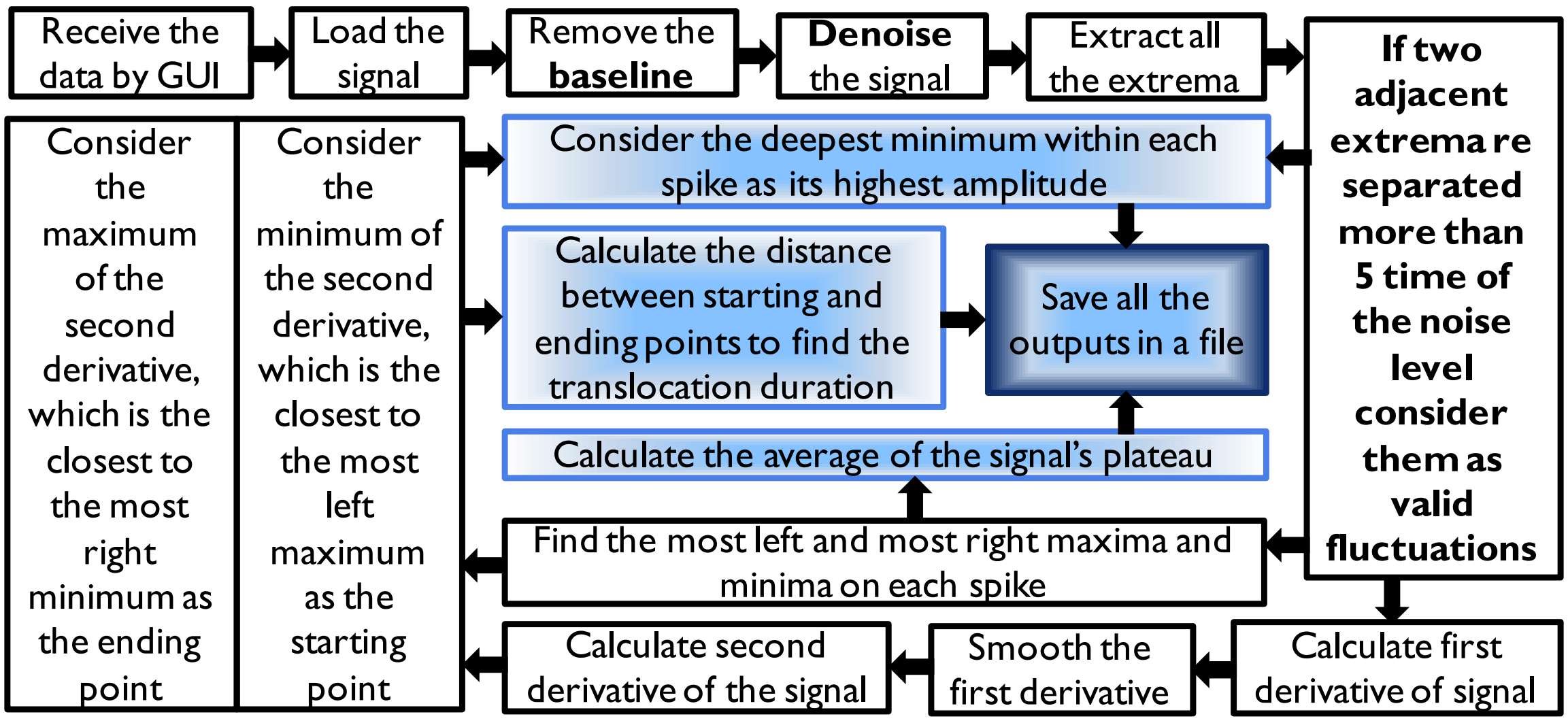

**Figure 7** *Block diagram the proposed software.*

## 2.5 Amplitude and Duration Measurement

Previously, there were a few slightly different approaches for measuring the spike's duration, all based on thresholding; the distance between the two adjacent crossing points of the threshold, the distance between the point that the signals leaves the baseline and the point that it returns to the baseline [6, 23, 24, 25], or considering the full width half max of the spike to reduce the effects of the limited bandwidth [2]. Pedone et al. have introduced the distance between the point at which the signal leaves the baseline and the last local minimum of the spike as the spike's duration [8]. Although such definitions may considerably reduce the effect of bandwidth, they have no physical meaning and are prone to noise within the spike. Thus, a new definition for the spike's duration is introduced, which is least affected by the measurement bandwidth and noise, and most importantly has a physical meaning. This definition considers the time that the center of nanoparticle enters the sensing zone as the start of the spike and the time that its center leaves the sensing zone as the ending point. A simple analytical calculation shows that these are the times at which the second derivation of the signal reaches its maximum and minimum. Since the measurement bandwidth only changes the rise time and fall time of the signal and maybe the peak amplitude of the signal, it is not affecting the spike's duration that has measured by this method (fig. 6.b).

Conventional algorithms are considering the distance between the average of the spike's plateau or its highest point, to the baseline as the spike's amplitude. In addition to these amplitudes, the proposed software also calculated the distance between the highest maximum and minimum within a spike as its amplitude. The later definition of the amplitude is not affected by the baseline drift and results in narrower distributions.